\documentclass[pra,floatfix,preprint,nofootinbib,eqsecnum,12pt]{revtex4}
\usepackage{epsfig}
\usepackage{bm}
\usepackage{amsmath}
\usepackage{setspace}
\input epsf
\numberwithin{equation}{section}
\renewcommand{\theequation}{\arabic{section}.\arabic{equation}}

\def\ed{ed.~by~}

\def\HP{$(I,\Psi)$}

\def\be{\begin{equation}}
\def\ee{\end{equation}}

\begin{document}
\vspace{1cm}

\title{How Nature  is Conformable to Herself: \\ A View from Quantum Cosmology\footnote{A pedagogical essay.}}
\author{James B.~Hartle}

\email{hartle@physics.ucsb.edu}

\affiliation{Santa Fe Institute, Santa Fe, NM 87501}
\affiliation{Department of Physics, University of California,Santa Barbara, CA 93106-9530}

\date{\today }

\begin{abstract}
\singlespacing
In his essay {\it Nature Conformable to Herself} the late Murray Gell-Mann expands on an observation of Newton that theories of seemingly disparate phenomena in the universe often make use of similar ideas and similar mathematical structure. Newton summarized that by saying that nature was very consonant and conformable to herself. This essay uses a model of quantum cosmology to illustrate how, why, and when nature is conformable to herself. 
\vskip .5in
\centerline{ \Large In memory  of Murray Gell-Mann, 1929-2019.\footnote{Murray Gell-Mann was the author's teacher and long time collaborator. We worked together for approximately thirty years to develop a formulation of quantum mechanics  that was general enough for cosmology --- decoherent histories  quantum mechanics.  This essay is dedicated to his memory, even knowing how much it would have been corrected, improved, argued with, clarified, its English improved, its spelling corrected, and seen in a wider context  were he available to read it. For a short appreciation of Murray Gell-Mann by the author see \cite{Har15b}.} } 
\end{abstract}




\maketitle

\bibliographystyle{unsrt}


\centerline{\Large  Nature will be very conformable to herself and very simple.} 
\centerline{(I. Newton, Opticks, Book III, part1, Query 31.) \cite{newtquote}}

\section{Introduction}
\label{intro1}
In his beautiful and deep essay {\it Nature Conformable to Herself: Arguments for a Unified Theory of the Universe} (abbreviated here by NCH) \cite{CONF,GM-Cajal}  the late Murray Gell-Mann
 described progress in fundamental physics as a progression of understanding  through `layers' of phenomena characterized by higher and higher energy scales and less and less experimental accessibility. He uses the metaphor `peeling away the skin of the onion' to characterize this progression. 

As he notes, the fundamental laws of nature are such that a rough similarity often applies in the effective theories that describe phenomena at these levels and, in particular,  similarities in the mathematical structures used. We typically do not have to invent new mathematics to enable the theories of the different levels. It is often already available or almost already there suggested by what we already have. This is an instance of `nature conformable to herself'. 

Beyond these connections, we note that many of the successful effective theories at the different levels have something of a universal character. They describe regularities that are exhibited  by  members of  various classes of subsystems of the Universe at a given level. An example is the regularities in luminosity exhibited by the various classes of stars.

But  what exactly defines these levels? {\it How} is  nature  is conformable to herself?  What is the origin of the universality sometimes displayed? If we are searching for a unified theory of the Universe as in Gell-Mann's subtitle these questions don't have definite answers.  Rather,  these ideas guide the search for such a theory. But if we have such a unified theory of the Universe, or even a sketch of one,  these questions should have derivable answers.  In this paper we are going to provide answers in a simple model of a unified theory of  the Universe in the context of quantum cosmology. Our aim is not to critique Gell-Mann's ideas in NCH but to illustrate and provide examples of them. 

At the present moment, unified theories of the Universe are thought to consist of two parts:  First, a theory of the Universe's quantum dynamics specified by an action $I$ on histories of spacetime geometry and matter fields.  Superstring theory is an example. 
Second, a theory of the Universe's quantum state $\Psi$ specified by an appropriate wave function such as the no-boundary wave function of the Universe \cite{NBWF}.  Roughly speaking the dynamical theory $I$ describes regularities in time while the quantum state $\Psi$ governs regularities in space.\footnote{ The NBWF exhibits some measure of unification of the theories of dynamics and state since  the dynamical action $I$ has an explicit role in its construction. This unification is also manifest in the holographic construction of the NBWF \cite{Holo}.}

For a simple model to discuss we assume a minisuperspace of homogeneous, isotropic, {\it spatially closed} cosmological geometries whose metrics are characterized by a scale factor $a$ viz. 
\be
\label{metrics}
ds^2=-dt^2+ a(t)^2 d\Omega_3^2
\ee
where $d\Omega_3^2$ is the metric on the unit 3-sphere.

For the quantum state of the Universe  $\Psi$ we assume the no-boundary wave function \cite{NBWF} in the minisuperspace context.  The NBWF can be thought of  as a cosmological analog of a ground state.   For the dynamics $I$  we assume Einstein gravity coupled to a  collection of quantum fields,  a useful example being a single scalar field $\phi(x)$ moving in a  potential $V(\phi)$. We take \HP\ to be a model of what Gell-Mann means by a  `unified theory of the Universe'' in the subtitle to his paper.  The  output  of this quantum theory of the Universe are quantum probabilities for histories of what goes on in the  Universe over time.

\section{Quantum Mechanics for the Universe} 
\label{qmu}
 Calculation of the quantum probabilities for histories of the spatially closed cosmologies  described above from \HP\ requires a quantum mechanics of a closed system where everything is inside that system. In particular, observers and what they observe along with any apparatus they use to make observations are physical systems within the  closed system not somehow outside it. Familiar textbook quantum mechanics thus has to be generalized to be used for cosmology (see e.g \cite{Har19}). We shall use the decoherent (or consistent) histories formulation of quantum mechanics. \cite{classicDH}.  Decoherent histories quantum theory (DH) is the work of many \cite{classicDH}. The formulation we use here is the work of Gell-Mann and the author \cite{GH1,GH2,GH3,GH4,GH5,GH6,GH7,GH8,GH9}. On many essential points it coincides with the independent earlier  consistent histories  formulation of quantum theory of Griffiths and Omn\`es \cite{classicDH}.  DH  can be seen as a generalization, clarification, and, to some extent, a completion of the of the program started by Everett \cite{Eve57}  for a quantum mechanics of a closed system like a closed Universe.  The  details of DH  will not be used this paper but a very brief summary is in the Appendix. 
 
 The outputs of the unified theory of the Universe \HP\ are probabilities for the individual members of decoherent  sets of coarse-grained alternative time histories of what goes on in the closed Universe --- a classical cosmological geometry, its expansion, the nucleosynthesis of the light elements, the growth of fluctuations away from homogeneity and isotropy, the formation of the cosmic background radiation, the nucleosynthesis of elements, the formation of galaxies, stars, planets, the evolution of their biota, the outcomes of measurements carried out by human observers, the outcomes of the experiments at CERN  etc. etc.  Model theories of the kind under discussion have been successful at predicting many of these features of our universe, e.g \cite{11,12,13,14,21,22,23}.
 
 It is important to emphasize that DH does not assume  many of properties of the Universe that are sometimes assumed in less general formulations of quantum theory.  DH does not assume a classical cosmological spacetime, or a  quasiclassical realm,  or separate classical and quantum worlds,  or observers,  or  apparatus,  or  measurements, or approximately isolated subsystems, nor does it assume the textbook  Copenhagen formulation of quantum theory. 
 These are all true features of our Universe that can be described in DH. However they  are not universal features.  Rather they  emerge  approximately and  at particular times and layers in its evolution e.g \cite{Har19}. 
 
 \section{Layers of the Onion }
 \label{layers}
In this model Gell-Mann's `layers' can be understood  as epochs in the Universe's history in which new phenomena emerge. We mean emergence in the sense Gell-Mann uses it in the essay NCH under discussion: ``You don't need something more to get something more''.  The layers are already there in \HP.  The theory just has to be looked at in the right way to bring them out. 

Once classical spacetime has emerged from the quantum gravitational layer at the beginning, Gell-Mann's notion of emergence can be given a quantitative meaning in cosmology that can be illustrated by the following example:  Choose an epoch in the history of the universe marked, say, by a time before the present in the classical history of spacetime geometry. Using DH, calculate the probability from \HP\ that that there were no stars before this epoch but many after it.  If that probability is high we say that stars emerged at that epoch.  However difficult the calculation of this probability might be we did not have to add anything to \HP\ and DH. 
Thus,  the question is not whether stars emerged at a given epoch or did not, there is rather a probability defined by \HP\ and DH that they emerged and a probability that they did not. Further the question of emergence of a given class of physical systems depends on what coarse graining is used to describe them.  For instance it might be probable that birds emerged at particular epochs but very improbable that a  particular species like carrier pigeons emerged.


A well known  set of layers for our universe can be characterized very roughly as follows starting with the big bang\footnote{Discussions can be found in many  books on the physics of the early Universe, e.g. \cite{KT89}.}.
 
 \begin{itemize}
 \item{\sl Quantum Spacetime:}  At the center of the onion  spacetime geometry does not behave classically but fluctuates quantum mechanically in an essential way.  We therefore need a theory of quantum gravity. Since gravity couples to all forms of energy we also need  theories of how it couples to all the matter degrees of freedom that will participate in the emergence of effective theories at lower layers. As Gell-Mann suggests in his essay. this could be string theory in one form or another.
 
 \item{\sl Classical Spacetime,  the Quasiclassical Realm,  and Copenhagen Quantum Mechanics:} Classical behavior is not a given in quantum theory. It is a matter of quantum probabilities. A system behaves classically when, in a suitably coarse-grained set of alternative histories,  the probabilities are high for those that exhibit correlations in time governed by classical deterministic laws. As the universe expands, and characteristic energies decrease, an approximately homogeneous and isotropic cosmological classical spacetime emerges.  Along with that  there emerges the wide range of time, place, and scale on which deterministic laws of classical physics apply  that constitute the quasiclassical realm of every day experience \cite{Har07b,GH7}.  
 
 Along with classical spacetime the familiar textbook  ``Copenhagen''  formulation of quantum mechanics of measurement situations emerges.  This was not possible at a higher level  because Copenhagen quantum mechanics assumes classical spacetime.  It assumes it just to define what the time $t$ means in the Schr\"odinger equation. Further, Copenhagen quantum mechanics requires a classical spacetime to define foliating families of spacelike surfaces through which the unitary evolution of the wave function proceeds. Plausibly, Copenhagen quantum mechanics emerges along with its prerequisite classical spacetime \cite{Har93c,Har19}.
 
 \item{\sl Elementary Particles:} 
 \label{elementary}
 The next layer is characterized by an approximately homogeneous and isotropic expanding classical spacetime containing a gas of  elementary particles such as protons, neutrons,  electrons, neutrinos, and photons interacting by strong, weak, and electromagnetic  forces. The gas rapidly comes to thermodynamic equilibrium characterized by a temperature $T$.  This layer is the domain of elementary particle physics. 
 
 \item{\sl Big Bang Nucleosynthesis:}
 \label{nucleosynthesis} 
 As the Universe expands its temperature $T$ drops. When it has dropped sufficiently far, protons and neutrons combine to synthesize some light elements  ---  $H, He,D, He^3, Li^7$.... etc.  This is a layer  governed primarily by  nuclear physics as the effective theory. 
 
 \item{\sl Recombination:}
 \label{recomb}
 As the Universe continues to expand its temperature continues to drop.  Nuclei and electrons combine to make atoms --- mostly of hydrogen and helium but with trace amounts of other elements. We are then in the regime of atomic physics.
 
 \item{\sl Large Scale Structure ---Galaxies, Stars, Planets, Black Holes:} 
 \label{lss}
 In the models considered in this paper the no boundary quantum state (NBWF) implies that  homogeneous and isotropic   classical spacetime of the early universe contained  small quantum fluctuations away from these symmetries \cite{HHaw85}.  Those fluctuations grew under the action of gravitational attraction to produce the large scale structure of the Universe that we see today in the distribution of the galaxies, stars, planets and black holes and in the tiny fluctuations in the temperature of the cosmic background radiation away from isotropy roughly 400,000 years after the big bang.

 \item{\sl Approximately Isolated Subsystems and  their Frozen Accidents:}
 \label{isolated}
 Early in the Universe there were no isolated subsystems. Today  our  Universe exhibits many different kinds of approximately isolated subsystems.:  galaxies, stars, planets, plants and animals on Earth, bacteria, observers, apparatus, tables,  chairs, human beings, you, me etc, etc.   Many classes of isolated subsystems exhibit regularities that are consistent with the basic laws of physics but do not follow from them as in the example in Section \ref{univreg} below.    Rather many of their regularities resulted from frozen accidents --- chance events that had long term and/or wide spread consequences \cite{QJ94}.
 The evolution of stars,  the evolution of the human genome \cite{FC68} and  the evolution of different bird species on Earth  are just  a few of examples of a great many emergent features of the Universe that are probably the results of frozen accidents. 
 
 \end{itemize}
 
 \section{Non-Surprises}
 \label{non-surprises}
 
 \subsection{Mathematics}
 \label{math}
 In this model it is not surprising that the mathematics useful at one level is related to that at another level. There is only one mathematics underlying the unified theory \HP.  Different parts of this mathematics may be more useful than others at different levels but they are all partial expressions of the same mathematical structure. The closer the mathematics at one level is to the mathematics of the next, the simpler it is to make connections between them, to augment them, or to invent any new mathematics needed.\footnote{The fact that related mathematical ideas apply in different layers for instance  might be taken as an example of  the `unreasonable effectiveness of mathematics in the natural sciences' \cite{Wig60}. But in fact the connections have a simple explanation in their origin in one  unified theory of the Universe.}
 
 \subsection{Universality}
 \label{universality}
 The theory \HP\ is universal in the elementary sense that it predicts probabilities for all possible decoherent  sets of coarse-grained alternative histories of the Universe.
 But it is also universal in two more specific senses:
 
 \subsubsection{Universal Regularities Obeyed by All Subsystems}
\label{univreg}
 If a cat, a cannonball, and an economics textbook are all dropped from the same height, they fall to the ground with the approximately the  same acceleration under the influence of gravity. The equality of the gravitational accelerations of different things is one of the most accurately tested laws of physics. The accelerations of the Earth and the Moon toward the Sun  are known to be equal to an accuracy of a few parts in a thousand billion.  That is an example of a universal regularity following from the  physics of the classical spacetime geometry that emerged shortly after the big bang.  
 
 \subsubsection{Regularities for Some Subsystems with a Common Origin} 
 \label{classes}
 The example above says very little about cats, cannonballs or  economics.  The class of cats for example has its own regularities that are not direct consequences of \HP.  Rather these mostly arise from the  frozen accidents of biological evolution as described in Section \ref{isolated}.
 
 For instance, galaxies arose from the collapse of quantum density fluctuations in the early Universe. The specific arrangement of the galaxies in the Universe is mostly an accident.  The origin mechanism of galaxies explains the similarities between them.  
 
 \section{Complexity and Simplicity}
 \label{complexity}
 The evidence of the observations is that the Universe was simpler earlier than it is now --- more homogeneous, more isotropic, more nearly in thermal equilibrium\footnote{We aim here at an intuitive assessment of complexity. The discussion could be made more quantitative by using an appropriate measure of complexity such as the effective complexity of Gell-Mann and Lloyd \cite{GL04}. This is roughly the length of the shortest description of the system's regularities. The cosmic background radiateion (CMB)  is simple because its regularity  is its isotropy which can be described succinctly.}.
 
 The Universe becomes more complex over time as more and more structure emerges like galaxies, stars, planets, etc, until  roughly the present epoch. Indeed, some level of complexity along with exploitable regularities  is necessary for information gathering and utilizing systems (IGUSes) like ourselves to exist
 \cite{Har16}. Were the Universe exactly homogeneous for example there would be no other subsystems for an IGUS like us to  exploit to function by obeying the evolutionary imperatives ``get food, yes'',  ``be food, no'', ``make more, yes''. 
 
 However, complexity doesn't grow forever. The accelerated expansion of the Universe driven by the cosmological constant means that ithe Universe will become cold dark, nearly empty, approximately  homogeneous, isotropic at late times  and, in short, in an intuitive sense, it will become  {\it simple}.
 
 \section{Going In and Coming Out (of the Onion)}
 \label{going} 
 Historically, the effective theories governing the physics  of the different layers were discovered  in  laboratory experiments that accessed higher and higher  energy scales (toward the center of the  onion). The rough sequence of discovery was: 
 classical physics (including curved spacetime),  atomic physics, nuclear physics, elementary particle physics. 
 
 Experimental progress upward resulted in the discovery of effective dynamical theories for the phenomena at each level. These are consistent in the sense that theories at a lower level can be derived from an appropriate approximation to  the theory at a higher level assuming classical spacetime. 
 
 It is remarkable that the successive effective theories that describe the layers  can  all be expressed in a form that is local in classical spacetime once that emerged from the quantum fog at the beginning.
 Thus, these  effective theories could be discovered and tested in localized laboratories like those at CERN and then extrapolated to cosmology.  It seems likely that this upward progress in experimental energy scales is what Gell-Mann had in mind  in peeling back  layers to move toward the center of the theoretical onion.
 
 We can speculate that the upward progress  in accessible experiment made possible by locality is the reason that candidates for the dynamical theory $I$ part of a unified theory were developed before theories of the quantum state of the Universe $\Psi$. Regularities in time were often discovered in laboratories on Earth in experiments probing  modest spatial scales.  Discovering and testing regularities in space on cosmological sales has required much more sophisticated (and expensive) technology only available more recently.
 
 Experimental progress upward in energy is limited.  It is unlikely that we will build  an accelerator that would directly probe the Planck scale energies that characterize many of today's candidates for unified theories. That not least because at those energies we would be in the regime where spacetime geometry itself is fluctuating quantum mechanically prohibiting a definition of locality.
 
 Rather we must think of the big bang as an experiment that was done $13.7$ billion years ago leaving  data that today is scattered over the approximately 14 billion light-year size of our cosmological horizon and then  use that data to understand the Universe at the innermost  layer of the theoretical onion
 
 \section{Conclusion}
 \label{conclusion}
 Towards the end of his essay (NCH) Gell-Mann  writes:  ``the conformability of nature to herself, the applicability of the criterion of simplicity, and the utility of certain parts of mathematics in describing physical reality ...  are thus emergent properties of the fundamental laws of physics''  --- presumably what he calls elsewhere ``a unified theory of the Universe''.

 In this paper we have presented a contemporary model of such a unified theory of the Universe --- a model quantum cosmology. The model is based on three theoretical inputs: a theory of quantum dynamics of the Universe $(I)$, a theory of its quantum state $(\Psi)$, and the principles of decoherent histories quantum mechanics (DH) for calculating probabilities for the individual members of decoherent sets of coarse-grained alternative histories of what happens in the universe.  
 
 In this paper we have shown  how the history of the Universe could be divided up into layers characterized by the emergence of new phenomena at new energy scales and times described by new effective theories. Links between the various levels derivable from \HP\  implement the idea of `conformable'. We showed how the mathematics useful at one level was naturally connected with that an another level because there was only one mathematical structure for implementing the unified theory of the Universe. We provided a quantitative measure of  the emergence in the layers based on \HP\ and the DH quantum mechanics of cosmological history. 
 
 Thus quantum cosmology provides both the context and the explanation of how nature is conformable to herself. The author believes that Gell-Mann would have agreed with this.


 
 \vskip .2in
 
\acknowledgments

The author of course acknowledges many conversations with Murray Gell-Mann on quantum mechanics and quantum cosmology. He thanks Thomas Hertog and Mark Srednicki for conversations on the quantum mechanics of the Universe and quantum cosmology  over a long period ot time.  This work was supported in part by the US National Science Foundation under grants PHY18-18018105, and PHY-1911701. 

	\renewcommand{\theequation}{\arabic{equation}}
 \appendix
 \section{Histories and Realms}
 \label{histories-realms}
 This appendix presents a bare bones description of how the theory \HP\  predicts probabilities for which of a decoherent  set of alternative coarse-grained histories happens in the Universe. Many more details and specific models can be found in e.g. \cite{classicDH,Har93a}. 
 
 Imagine a closed quantum system of a particle in a box.
 The simplest notion of a set of histories is described by specifying a sequence of  yes-no alternatives at a series of times. 
For example: Is a particle in this spatial region $R$ of the box at this time --- yes or no?  This alternative is {\it coarse-grained}  because it does not  follow the particle's position exactly but only whether it is in $R$ or not. The alternative `yes' is represented by the projection operator $P_R$ on the region $R$ and `no' by $I-P_R$.  More generally coarse-grained yes-no  alternatives at one moment of time are described by an exhaustive set of exclusive Heisenberg picture projection operators $\{P_\alpha(t)\}$, $\alpha=1,2,3,\cdots$ acting on the Hilbert space of the particle.   These satisfy:
\be
\label{projections1} 
P_\alpha(t)P_{\alpha'}(t) = \delta_{\alpha\alpha'} P_\alpha(t), \quad \sum_\alpha P_\alpha(t) = I . 
\ee

Projections on bigger subspaces are more coarse grained, projectors on smaller subspaces are finer grained. 

Projection operators representing the same quantity at different times are connected by unitary evolution by the Hamiltonian $H$ when there is one, viz.  
\be
P_\alpha(t') = e^{iH(t'-t)}P_\alpha(t) e^{-iH(t'-t)} . 
\label{un-evol}
\ee
For example, to describe the quasiclasscal realm in a fixed background spacetime containing many particles  the projections would be products of projections  for each subvolume onto ranges of values of the quasiclassical variables  --- averages over the subvolumes  of energy, momentum, and number \cite{GH7,GH9}.

A set of alternative coarse-grained histories is specified by a sequence of such sets of orthogonal projection operators at a series of times $t_1,t_2, \cdots t_n$. An individual history corresponds to a particular sequence of events  $\alpha \equiv (\alpha_1,\alpha_2, \cdots, \alpha_n)$ and is represented by the corresponding chain of projections:
\be
C_\alpha \equiv P^n_{\alpha_n}(t_n) \dots P^1_{\alpha_1}(t_1) .
\label{class-closed}
\ee
Branch state vectors corresponding to the individual histories can be defined by
\be
\label{branch}
|\Psi_\alpha\rangle \equiv  C_\alpha |\Psi\rangle . 
\ee

The set of histories (medium) decoheres when the quantum interference between branches is negligible
\be
\label{decoherence}
\langle\Psi_\alpha|\Psi_\beta\rangle  \approx  0, \alpha \ne \beta
\ee
As a consequence of decoherence probabilities $p(\alpha)$  consistent with usual probability rules can be assigned to the individual histories that are 
\be
\label{probs}
p(\alpha) = |||\Psi_\alpha\rangle||^2 = ||C_\alpha |\Psi\rangle||^2.
\ee
A decoherent set of alternative coarse-grained histories is called a `realm' for short.  

This description of histories may seem similar to those for sequences of ideal measurements in Copenhagen quantum mechanics.  However, there are at least three crucial differences. First, there is no posited separate classical world as in the Copenhagen formulation. It's all quantum. Second, the alternatives represented by the $P$'s are not restricted to measurement outcomes. They might, for example, refer to the orbit to the Moon when no one is looking at it, or to the magnitude of density  fluctuations in the early Universe when there were neither observers nor apparatus to measure them. Laboratory measurements can of course be described in terms of correlations between two particular kinds of  subsystems of the Universe --- one being measured, the other doing the measuring. But laboratory measurements play no central role in formulating DH, and are just a small part of what it can predict Indeed, the Copenhagen quantum mechanics of measured subsystems is an approximation appropriate for measurement situations to the more general quantum mechanics of closed systems. See, e.g. Section II.10 of \cite{Har91a}. Third, the $P$'s can refer to  alternatives describing observers and other IGUSes and to histories of how they formed and what they are doing:  for instance whether they are making measurements or not.

\vskip .2in


\noindent{\bf References \cite{GH1,GH2,GH3,GH4,GH5,GH6,GH7,GH8,GH9} are the papers on  quantum mechanics written jointly by Murray Gell-Mann and the author.}


\begin{thebibliography}{99}
\bibitem{newtquote}  The translated quote is from the Dover edition of Opticks, p. 397.
\end{thebibliography}
\end{document}